\documentclass[useAMS,usenatbib]{mn2e}
\usepackage{txfonts}
\usepackage{natbib}
\usepackage[all]{xy}
\usepackage[dvips]{graphicx}

\def\del#1{{}}

\newcommand{\bra}{\langle}
\newcommand{\ket}{\rangle}

\newcommand{\dd}{\mathrm{d}}

\newcommand{\p}{\mathrm{p}}

\newcommand{\trace}{\mathrm{tr}}

\newcommand{\dirac}{\delta_D}
\newcommand{\id}{\mathrm{id}}

\title[tomography with polynomials]
{Weak lensing tomography with orthogonal polynomials}
\author[B.M. Sch{\"a}fer, L. Heisenberg]
{Bj{\"o}rn Malte Sch\"afer\thanks{e-mail: bjoern.malte.schaefer@uni-heidelberg.de} and Lavinia Heisenberg$^2$\\
$^1$ Astronomisches Recheninstitut, Zentrum f{\"u}r Astronomie, Universit{\"a}t Heidelberg, M{\"o}nchhofstra{\ss}e 12, 69120 Heidelberg, Germany\\
$^2$ D{\'e}partement de Physique Th{\'e}orique, Centre for Astroparticle Physics, Universit{\'e} de Gen{\`e}ve, 24, quai Ernest Ansermet, 1211 Gen{\`e}ve, Switzerland}

\begin{document}
\pagerange{\pageref{firstpage}--\pageref{lastpage}}
\pubyear{2011}
\maketitle
\label{firstpage}

% --- abstract --- %
\begin{abstract}
The topic of this article is weak cosmic shear tomography where the line of sight-weighting is carried out with a set of specifically constructed orthogonal polynomials, dubbed TaRDiS (Tomography with orthogonAl Radial Distance polynomIal Systems). We investigate the properties of these polynomials and employ weak convergence spectra, which have been obtained by weighting with these polynomials, for the estimation of cosmological parameters. We quantify their power in constraining parameters in a Fisher-matrix technique and demonstrate how each polynomial projects out statistically independent information, and how the combination of multiple polynomials lifts degeneracies. The assumption of a reference cosmology is needed for the construction of the polynomials, and as a last point we investigate how errors in the construction with a wrong cosmological model propagate to misestimates in cosmological parameters. TaRDiS performs on a similar level as traditional tomographic methods and some key features of tomography are made easier to understand.
\end{abstract}

% --- keywords --- %

% --- section: introduction --- %
\section{Introduction}
Weak gravitational lensing by the cosmic large-scale structure \citep{1991MNRAS.251..600B, 1992grle.book.....S, 1994CQGra..11.2345S, 1994A&A...287..349S, 1998MNRAS.301.1064K} is regarded as a very promising way for investigating the properties of the cosmic matter distribution and the measurement of cosmological parameters. The primary tool is the angular spectra of the weak lensing convergence \citep{1997ApJ...484..560J, 1999ApJ...514L..65H, 2001ApJ...554...67H, 2004PhRvD..70d3009H}, which has, in the context of dark energy cosmologies with adiabatic fluctuations in the dark matter distribution, the potential to deliver parameter constraints competitive with those from the cosmic microwave background.

The weak cosmic shearing effect of the large-scale structure has been detected by four independent research groups a decade ago \citep{2000A&A...358...30V, 2000astro.ph..3338K, 2000MNRAS.318..625B, 2000Natur.405..143W} and now parameter constraints derived from weak lensing spectra coincide well with those from the CMB, with a small tension concerning the parameters $\Omega_m$ and $\sigma_8$. \citep{2009A&A...497..677K, 2010MNRAS.405.2381K}. These topics are reviewed in detail by \citet{2001PhR...340..291B} and \citet{2010RvMP...82..331B, 2010CQGra..27w3001B}.

Together with the property of weak lensing to map out fluctuations in the cosmic matter distribution in a linear way, tomographic methods have been shown to offer superior precision in the determination of cosmological parameters \citep{1999ApJ...522L..21H, 2002PhRvD..66h3515H, 2002PhRvD..65b3003H, 2004ApJ...601L...1T, 2006JCAP...06..025H}, in particular concerning the properties of dark energy \citep{2001PhRvD..64l3527H, 2002PhRvD..65f3001H, 2009JCAP...04..012H, 2010GReGr..42.2177H, 2011arXiv1105.0993L, 2011MNRAS.413.1505A}. The idea of tomography is a division of the galaxy sample used for shear estimation into a number of bins in distance. This splitting is the reason for two important advantages of tomography: The weak lensing convergence is a line of sight-integrated quantity of which one measures the angular fluctuation statistics. Angular convergence spectra, however, convey less information compared to the fluctuations statistics of the three-dimensional matter distribution because in the projection process, a mixing of scales is taking place. Secondly, cosmological parameters can have different influences on the amplitude of the weak lensing convergence at different redshifts, which would be averaged out. Combining  convergence spectra measured for each subset of galaxies are able to alleviate these these two problems. Related developments include shear ratio measurements, where the same tidal fields are measured with galaxy populations at different redshifts \citep{2003PhRvL..91n1302J, 2007MNRAS.374.1377T}, cross correlation cosmography, where part of the galaxies are used for constructing a template on which one observes the shearing effect of more distant galaxies \citep{2004ApJ...600...17B}, and finally three-dimensional weak shear methods, which provide a direct reconstruction of the fluctuations of the cosmic density field \citep{2005PhRvD..72b3516C, 2006MNRAS.373..105H, 2011MNRAS.413.2923K}. Common to these methods are nonzero covariances between measurements of different redshifts. This is a natural consequence that lensing is caused by the large-scale structure between the lensed objects and the observer, so that the light from different galaxy samples has to transverse partially the same large-scale structure for reaching the observer.

Our motivation is to revisit weak lensing tomography and to construct a weighing scheme for the galaxies as a function on distance, such that the covariances between spectra estimated from differently weighted galaxy samples has a particular simple, diagonal shape. For this goal, we construct a set of orthogonal polynomials in distance, which provides a diagonalisation of the covariance, and which, when employed in the derivation of the convergence spectrum, projects out statistically independent information about the large-scale structure.

After a brief summary of key formul{\ae} for cosmology, structure formation and weak lensing in Sect.~\ref{sect_cosmology}, we describe the construction of TaRDiS-polynomials and investigate their properties in Sect.~\ref{sect_tomography}. Their statistical and systematical errors are evaluated in Sects.~\ref{sect_statistics} and~\ref{sect_systematics}, respectively, and we summarise our results in Sect.~\ref{sect_summary}. The reference cosmological model used is a spatially flat $w$CDM cosmology with Gaussian adiabatic initial perturbations in the cold dark matter density field. The specific parameter choices are $\Omega_m = 0.25$, $\sigma_8 = 0.8$, $H_0=100\: h\:\mathrm{km}/\mathrm{s}/\mathrm{Mpc}$, with $h=0.72$, $n_s = 1$ and $\Omega_b=0.04$. The dark energy equation of state is set to $w=-0.9$ and the sound speed is equal to the speed of light ($c_s=c$) such that there is no clustering in the dark energy fluid.

% --- section: cosmology ---%
\section{cosmology and weak lensing}\label{sect_cosmology}

% --- subsection: dark energy cosmologies --- %
\subsection{Dark energy cosmologies}
In spatially flat dark energy cosmologies with the matter density parameter $\Omega_m$, the Hubble function $H(a)=\dd\ln a/\dd t$ is given by
\begin{equation}
\frac{H^2(a)}{H_0^2} = 
\frac{\Omega_m}{a^{3}} + (1-\Omega_m)\exp\left(3\int_a^1\dd\ln a\:\left[1+w(a)\right]\right),
\end{equation}
with the dark energy equation of state $w(a)$, for which we use the common parameterisation \citep{2001IJMPD..10..213C}
\begin{equation}
w(a) = w_0 + (1-a) w_a,
\end{equation}
where $w_0=-1$ and $w_a=0$ would correspond to $\Lambda$CDM. Comoving distance $\chi$ and scale factor $a$ are related by
\begin{equation}
\chi = c\int_a^1\:\frac{\dd a}{a^2 H(a)},
\end{equation}
such that the comoving distance is given in units of the Hubble distance $\chi_H=c/H_0$.

% --- subsection: CDM power spectrum --- %
\subsection{CDM power spectrum}
The linear CDM density power spectrum $P(k)$ describes statistically homogeneous Gaussian fluctuations of the density field $\delta$,
\begin{equation}
\bra\delta(\bmath{k})\delta(\bmath{k}^\prime)^*\ket=(2\pi)^3\dirac(\bmath{k}-\bmath{k}^\prime)P(k).
\end{equation}
It is composed from a scale invariant term $\propto k^{n_s}$ and the transfer function $T(k)$,
\begin{equation}
P(k)\propto k^{n_s}T^2(k).
\end{equation}
In low-$\Omega_m$ cosmologies $T(k)$ is approximated with the fit proposed by \citet{1986ApJ...304...15B},
\begin{equation}
T(q) = \frac{\ln(1+2.34q)}{2.34q}\left(1+3.89q+(16.1q)^2+(5.46q)^3+(6.71q)^4\right)^{-\frac{1}{4}},
\label{eqn_cdm_transfer}
\end{equation}
where the wave vector $k=q\Gamma$ is given in units of the shape parameter $\Gamma$. $\Gamma$ determines the peak shape of the CDM spectrum and is to first order given by $\Gamma=\Omega_mh$, with small corrections due to the baryon density $\Omega_b$ \citep{1995ApJS..100..281S},
\begin{equation}
\Gamma=\Omega_m h\exp\left(-\Omega_b\left(1+\frac{\sqrt{2h}}{\Omega_m}\right)\right).
\end{equation}
The spectrum $P(k)$ is normalised to the variance $\sigma_8$ on the scale $R=8~\mathrm{Mpc}/h$,
\begin{equation}
\sigma^2_R = \int\frac{\dd\ln k}{2\pi^2}\:k^3 P(k) W_R^2(k)
\end{equation}
with a Fourier transformed spherical top hat filter function, $W_R(k)=3j_1(kR)/(kR)$. $j_\ell(x)$ is the spherical Bessel function of the first kind of order $\ell$ \citep{1972hmf..book.....A}.

% --- subsection: structure growth --- %
\subsection{Structure growth with clustering dark energy}
Linear homogeneous growth of the density field, $\delta(\bmath{x},a)=D_+(a)\delta(\bmath{x},a=1)$, is described by the growth function $D_+(a)$, which is the solution to the growth equation \citep{1997PhRvD..56.4439T, 1998ApJ...508..483W, 2003MNRAS.346..573L},
\begin{equation}
\frac{\dd^2}{\dd a^2}D_+(a) + \frac{1}{a}\left(3+\frac{\dd\ln H}{\dd\ln a}\right)\frac{\dd}{\dd a}D_+(a) = 
\frac{3}{2a^2}\Omega_m(a) D_+(a).
\label{eqn_growth}
\end{equation}
Nonlinear structure formation enhances the CDM-spectrum $P(k)$ on small scales by a factor of $\simeq 40$, which is described by the fit suggested by \citet{2003MNRAS.341.1311S} which is gauged to $n$-body simulations of cosmic structure formation.

% ---  --- %
\subsection{Weak gravitational lensing}
The weak lensing convergence $\kappa$ provides a weighted line-of-sight average of the matter density $\delta$ \citep[for reviews, see][]{2001PhR...340..291B, 2010CQGra..27w3001B}
\begin{equation}
\kappa = \int_0^{\chi_H}\dd\chi\: W_\kappa(\chi)\delta,
\end{equation}
with the weak lensing efficiency $W_\kappa(\chi)$ as the weighting function,
\begin{equation}
W_\kappa(\chi) = \frac{3\Omega_m}{2\chi_H^2}\frac{D_+}{a}G(\chi)\chi,
\mathrm{~with~}
G(\chi) = 
\int_\chi^{\chi_H}\dd\chi^\prime\:q(z)\frac{\dd z}{\dd\chi^\prime}\frac{\chi^\prime-\chi}{\chi^\prime}.
\end{equation}
$q(z)$ denotes the redshift distribution of the lensed background galaxies,
\begin{equation}
q(z) = q_0\left(\frac{z}{z_0}\right)^2\exp\left(-\left(\frac{z}{z_0}\right)^\beta\right)\dd z
\quad\mathrm{with}\quad \frac{1}{q_0}=\frac{z_0}{\beta}\Gamma\left(\frac{3}{\beta}\right),
\end{equation}
and is the approximate forcasted distribution of the EUCLID mission \citep{2007MNRAS.381.1018A}. $z_0$ has been chosen to be $\simeq0.64$ such that the median of the redshift distribution is 0.9. With these definitions, one can carry out a Limber-projection \citep{1954ApJ...119..655L} of the weak lensing convergence for obtaining the angular convergence spectrum $C_\kappa(\ell)$, 
\begin{equation}
C_\kappa(\ell) = \int_0^{\chi_H}\frac{\dd\chi}{\chi^2}\:W_\kappa^2(\chi) P(k=\ell/\chi),
\end{equation}
which describes the fluctuation statistics of the convergence field.

% --- section:  --- %
\section{tomography with orthogonal polynomials}\label{sect_tomography}

% ---  --- %
\subsection{motivation for tomography}
Weak cosmic shear provides a measurement of the weighted, line of sight integrated tidal shears althoug it is more convenient to work in terms of the weak lensing convergence and the cosmic density field, which have statistically equivalent properties. Weak lensing therefore provides an integrated measurement of the evolution of the cosmic density field weighted with the lensing efficiency function. Being a line of sight-averaged quantity, the weak lensing convergence is statistically not as constraining as the full 3-dimensional density field, which is caused by the mixing of spatial scales in the Limber-projection. Additionally, parameters which enter the model in a nonlinear way can cause different effects on the weak lensing signal at different redshifts. This sensitivity would be averaged out in a long-baseline line of sight integration and the information would be lost.

The solution to this problem are tomographic methods: splitting up the lensing signal from different distances allows a larger signal strength and a higher sensitivity with respect to cosmological parameters because the line of sight-averaging effect is reduced. This comes at the cost of a more complicated covariance between the spectra measured, and a higher shape noise for each of the spectra, because the effective number of galaxies is reduced. There are different methods for exploiting evolution of the cosmic density field and the lensing sensitivity along the line of sight. Since the inception of tomography for investigating weak lensing convergence spectra \citep{1999ApJ...522L..21H, 2002PhRvD..65b3003H} and bispectra \citep{2004MNRAS.348..897T}, there have been many developments leading ultimately to cross-correlation cosmography \citep{2004ApJ...600...17B}, where a lensing template is derived from data and used to predict the weak shear signal on background sources, and to 3-dimensional weak shear methods, which are an unbinned, direct mapping of the cosmological density field \citep{2005PhRvD..72b3516C,2003MNRAS.343.1327H,2011MNRAS.413.2923K}. 

Common to these methods are strong covariances between spectra, which we aim to avoid by employing a set of specifically designed polynomials, which perform a weighting of the observed ellipticities in their distance, in a way that convergence cross-spectra computed with two different polynomials vanishes. This ensures that every polynomial projects out information about the deflection field which is statistically independent, and generates a covariance matrix of a particular simple shape. In a sense, TaRDiS is using a non-local binning of the ellipticities motivated by the anticipated weak lensing signal, which corresponds to the non-local nature of weak lensing.

% ---  --- %
\subsection{construction of orthogonal sets of polynomials}
We introduce a weighting of the distance distribution of the galaxies $q(\chi)=q(z)\dd z/\dd\chi = q(z)H(z)$ with a function $p_i(\chi)$, such that the lensing efficiency function reads
\begin{equation}
W_i(\chi) = \frac{3\Omega_m}{2\chi_H^2}\frac{D_+}{a}G_i(\chi)\chi
\end{equation}
with
\begin{equation}
G_i(\chi) = \int_\chi^{\chi_H}\dd\chi^\prime q(\chi)\:p_i(\chi)\:\frac{\chi^\prime-\chi}{\chi^\prime}.
\end{equation}
A line of sight-weighting of the weak lensing convergence with a function $p_i(\chi)$ yields the weighted convergence field $\kappa_i$,
\begin{equation}
\kappa_i = \int_0^{\chi_H}\dd\chi\:W_i(\chi)\delta,
\end{equation}
from which one obtains for the variance between two Fourier-modes $\kappa_i(\ell)$, 
\begin{equation}
\bra \kappa_i(\ell)\kappa_j^*(\ell^\prime)\ket = (2\pi)^2\delta_D(\ell-\ell^\prime) S_{ij}(\ell),
\end{equation}
with the weighted convergence spectrum $S_{ij}(\ell)$ for homogeneous and isotropic fluctuations of $\kappa_i$,
\begin{equation}
S_{ij}(\ell) = 
\int_0^{\chi_H}\frac{\dd\chi}{\chi^2}\:W_i(\chi)W_j(\chi)\: P(k=\ell/\chi).
\end{equation}
This expression provides the definition of a scalar product between the weighting functions $p_i(\chi)$ and $p_j(\chi)$ contained in $W_i(\chi)$ and $W_j(\chi)$, respectively,
\begin{equation}
\bra p_i,p_j \ket \equiv S_{ij}(\ell),
\label{eqn_scalar}
\end{equation}
which exhibits the necessary properties of being positive definite ($\bra p_i,p_i\ket\geq0$ and $\bra p_i,p_i\ket=0\leftrightarrow p_i\equiv0$), symmetric ($\bra p_i,p_j\ket=\bra p_j,p_i\ket$) and linear in both arguments. Starting point for the construction of orthogonal polynomials which would diagonalise $S_{ij}(\ell)$, $S_{ij}(\ell)\propto\delta_{ij}$, is the family monomials
\begin{equation}
p_i^\prime(\chi) = \left(\frac{\chi}{\chi_\mathrm{node}}\right)^i,
\end{equation}
where $\chi_\mathrm{node}$ can be chosen for placing the node of the first polynomial in a sensible way, in our case it is set to the median value of the galaxy redshift distribution, converted into comoving distance. The monomials $p_i^\prime(\chi)$ are subjected to a Gram-Schmidt orthogonalisation procedure, with the initial condition
\begin{equation}
p_0(\chi) = p_0^\prime(\chi) \equiv 1
\end{equation}
and, iteratively,
\begin{equation}
p_i(\chi) = p_i^\prime(\chi) - \sum_{j=0}^{i-1} \frac{\bra p_i^\prime,p_j\ket}{\bra p_j,p_j\ket}p_j(\chi).
\label{eqn_gram_schmidt}
\end{equation}
It should be emphasised that the orthogonalisation procedure needs to be repeated for every multipole $\ell$, and that we have omitted an additional index $\ell$ of the polynomials $p_i(\chi)$ for clarity. For illustration the polynomials are normalised using the norm induced by the scalar product defined in eqn.~(\ref{eqn_scalar}),
\begin{equation}
p_i(\chi)\leftarrow \frac{p_i(\chi)}{\sqrt{\bra p_i,p_i\ket}},
\label{eqn_norm_scalar}
\end{equation}
which appears in the denominator of eqn.~(\ref{eqn_gram_schmidt}) in a natural way. Clearly, the scalar product $S_{ij}(\ell)$ is equal to the convergence spectrum $C_\kappa(\ell)$ for $i=j=0$, in which case $p_0\equiv 1$.

% ---  --- %
\subsection{properties of orthogonal polynomials}
\begin{figure}
\resizebox{\hsize}{!}{\includegraphics{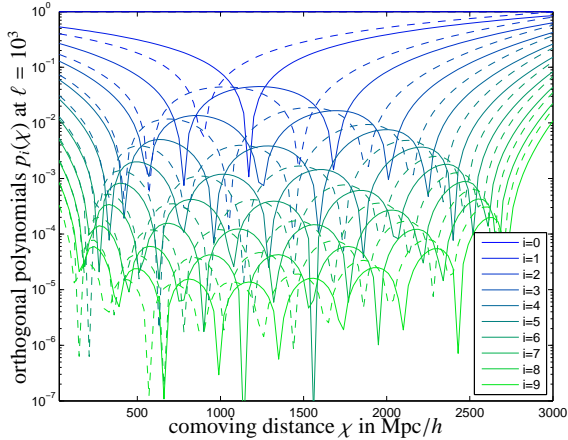}}
\caption{Orthogonal polynomials $p_i(\chi)$, $i=0\ldots9$, as a function of comoving distance $\chi$ constructed with the Gram-Schmidt algorithm for the weak lensing spectrum at $\ell=10^3$, with the lowest order polynomials on top, for both a linear (solid line) and a nonlinear (dashed line) CDM spectrum $P(k)$.}
\label{fig_polynomials}
\end{figure}

Fig.~\ref{fig_polynomials} shows the polynomials $p_i(\chi)$ at a fixed multipole order of $\ell=1000$. They exhibit a sequence of oscillations roughly at the positions where the previous polynomial assume their maximal values. When computing the polynomials for the nonlinear CDM spectrum instead for the linear one, the sequence of oscillations is shifted to smaller comoving distances. The reason for this shift is the larger amplitude of the nonlinear spectrum $P(k)$ for small $k=\ell/chi$, which causes a larger contribution of the weak lensing spectrum to be generated at high $\chi$.

\begin{figure}
\resizebox{\hsize}{!}{\includegraphics{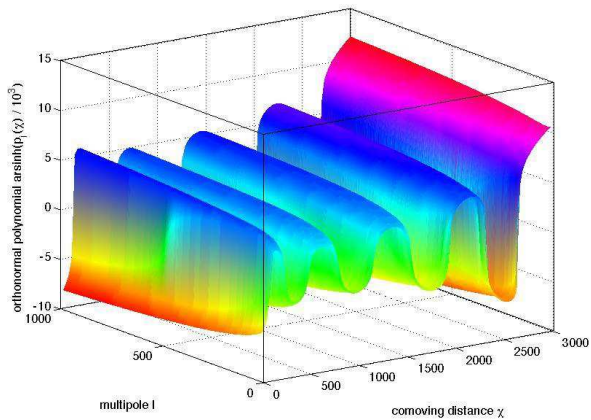}}
\caption{Orthonormal polynomials $p_i(\chi)$ as a function of multipole order $\ell$ and comoving distance $\chi$. The order of the polynomial has been fixed to $i=9$ and was constructed for a nonlinear CDM spectrum. One can see the slow variation of the $p_i(\chi)$ polynomials with multipole order $\ell$.}
\label{fig_polynomials_3d}
\end{figure}

The variation of the polynomials $p_i(\chi)$ with distance $\chi$ and multipole order $\ell$ is depicted in Fig.~\ref{fig_polynomials_3d}, in this case for the polynomial $i=9$. Again, small variations with multipole order $\ell$ are present, and the oscillations when varying $\chi$ are clearly seen.

\begin{figure}
\begin{tabular}{cc}
\resizebox{0.45\hsize}{!}{\includegraphics{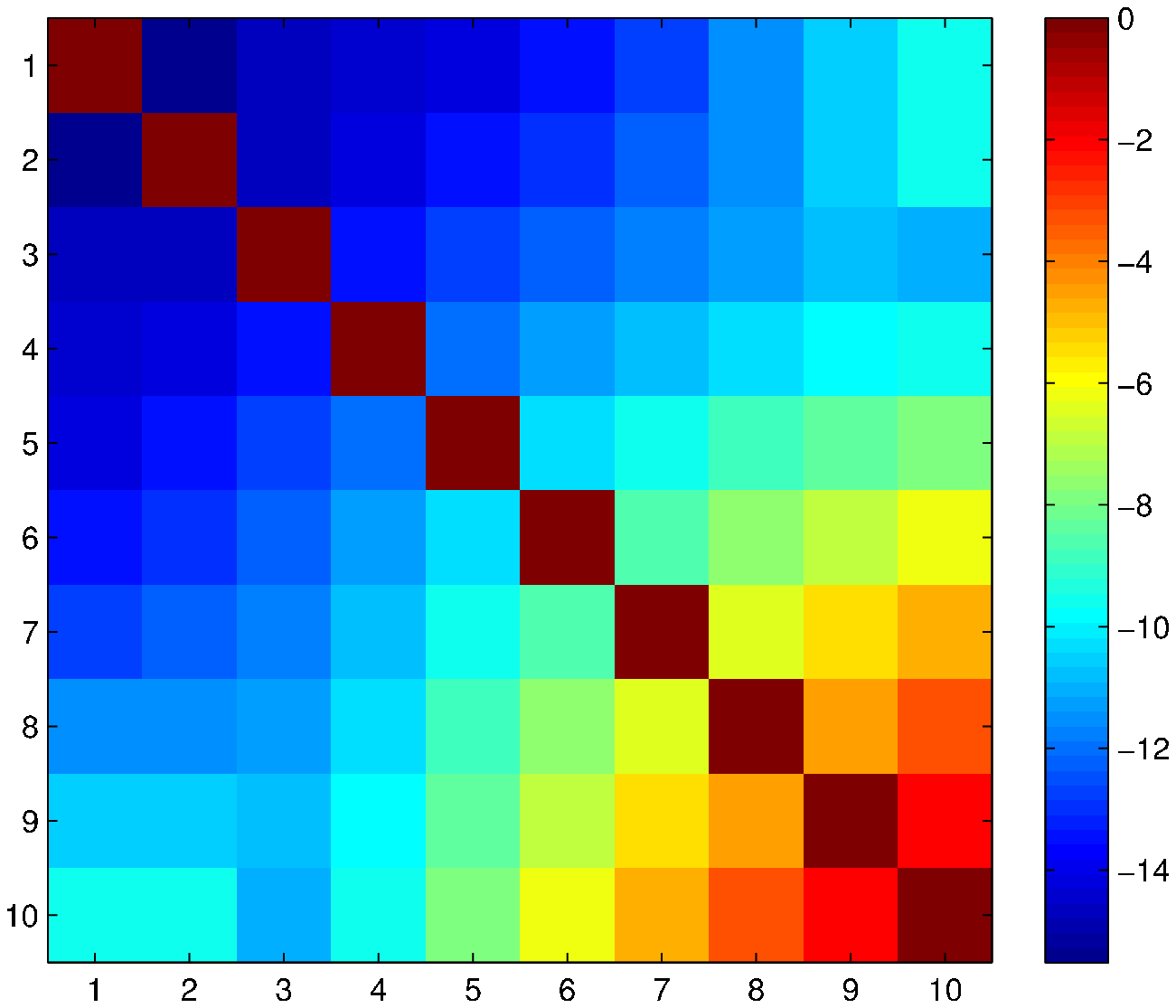}} &
\resizebox{0.45\hsize}{!}{\includegraphics{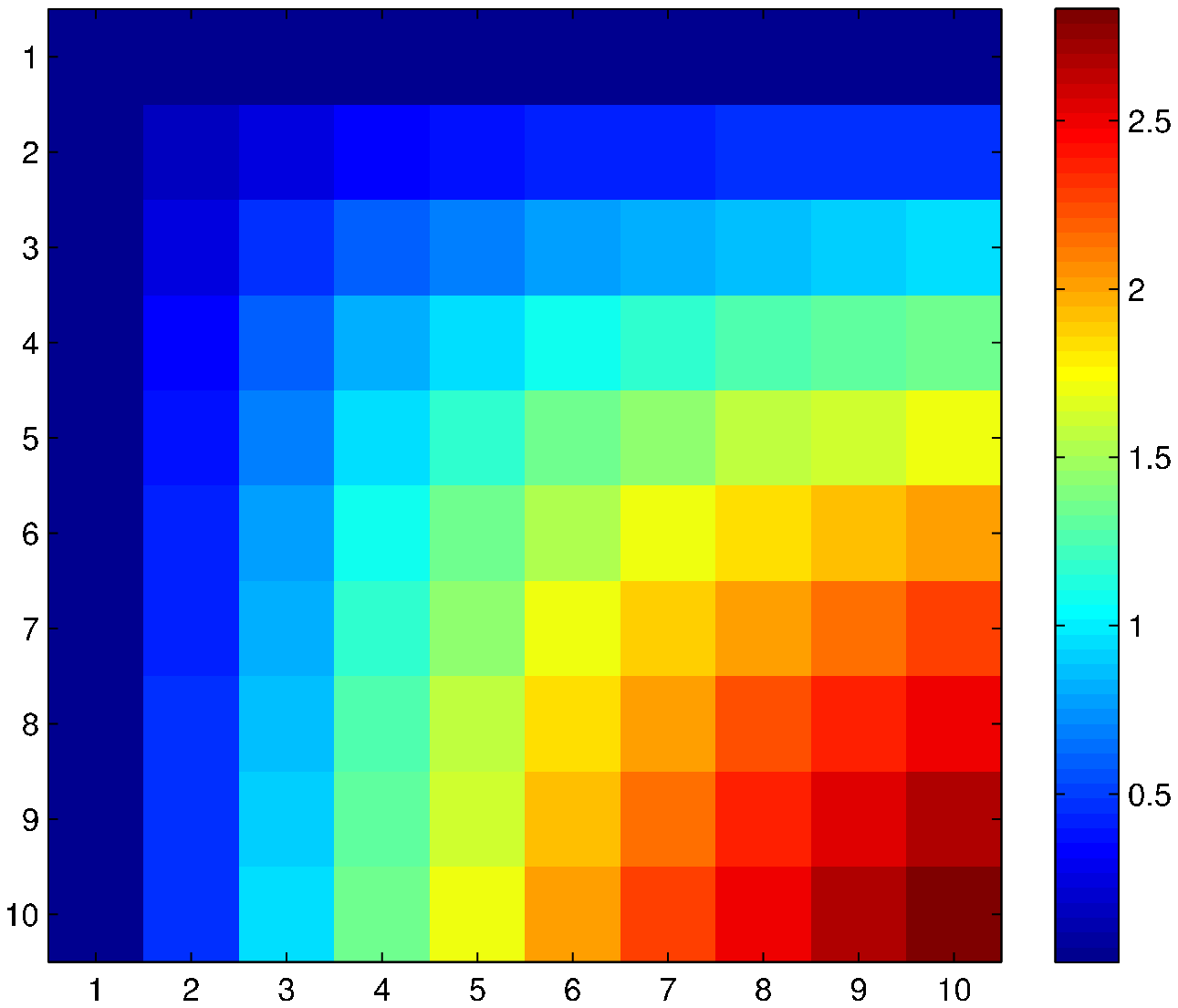}} 
\end{tabular}
\caption{Orthonormality relation for the polynomials $p_i(\chi)$, at $\ell=10^3$ in logarithmic representation, for the signal covariance $S_{ij}$ (left panel) and the noise covariance $N_{ij}$ (right panel), for the EUCLID survey characteristics, in units of the shape noise $\sigma_\epsilon^2/\bar{n}$.}
\label{fig_orthogonality}
\end{figure}

The orthonormality relation $\bra p_i,p_j\ket$ of of the polynomials $p_i(\chi)$ for the lensing signal as well as for the galaxy shape noise is given in Fig.~\ref{fig_orthogonality}. The polynomials are constructed to be orthogonal by the Gram-Schmidt procedure, but numerical noise is collected in the iterative process, such that the orthogonality relation is better fulfilled at small $i$ compared to larger $i$. Deviations from $\bra p_i,p_j\ket=0$ for $i\neq j$ are of the order of $\sim10^{-15}$ for low-order polynomials, but increase to values of $\sim10^{-4}$ at high order. This deterioration in orthogonality is a known drawback of the Gram-Schmidt procedure, in particular when dealing with sets of functions instead sets of vectors, which means that there is larger numerical noise in the evaluation of the scalar products, but is acceptable for the number of basis polynomials we are using in this work. One can already notice the main difference between the different approaches in tomography: Whereas the noise covariance would be diagonal in classic tomography with a very complicated structure of the signal covariance, in our case the shapes of the signal and noise covariance matrix are interchanged.

\begin{figure}
\resizebox{\hsize}{!}{\includegraphics{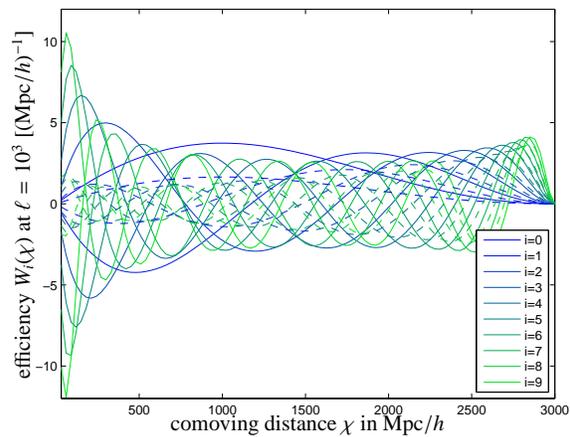}}
\caption{Lensing efficiency functions $W_i(\chi)$, $i=0\ldots9$, as a function of comoving distance $\chi$, for both linear (solid line) and nonlinear (dashed line) CDM spectra. The lensing efficiency functions have been constructed for $\ell=10^3$.}
\label{fig_lefficiency}
\end{figure}

Fig.~\ref{fig_lefficiency} gives an impression of the lensing efficiency function $W_i(\chi)$ modified by the polynomials $p_i(\chi)$ on an angular scale of $\ell=1000$, in comparison to that of weak shear without tomography, $W_0(\chi)\equiv W_\kappa(\chi)$ for $i=0$. It is quite interesting to see how the seemingly messy functions $W_i(\chi)$ disentangle and approach zero at large distances.

\begin{figure}
\resizebox{\hsize}{!}{\includegraphics{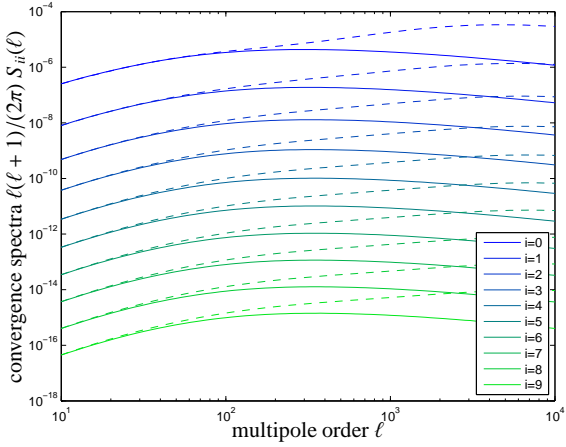}}
\caption{Weak lensing spectra $S_{ii}(\ell)$, $i=0\ldots9$, weighted with orthogonal polynomials $p_i(\chi)$, as a function of multipole order $\ell$, for both a linear (solid line) and a nonlinear (dashed line) CDM-spectrum $P(k)$. The spectra decrease in amplitude with increasing polynome order $i$ for unnormalised polynomials.}
\label{fig_spectra}
\end{figure}

Finally, polynomial-weighted weak lensing spectra $S_{ii}(\ell) = \bra p_i,p_j\ket$ are shown in Fig.~\ref{fig_spectra}. The spectra drop in amplitude, which is mostly an effect of the absence of normalisation, and there are in fact differences in shape, when higher-order polynomials are used (compare Fig.~\ref{fig_relative_spectra} in the appendix, where all spectra are scaled such that they assume the same value at a certain multipole.). By construction, these spectra provide statistically independent information of the cosmic large-scale structure. In the next sections, we will investigate statistical bounds and systematical errors on cosmological parameters, when the information from different spectra is combined.

% --- section:  --- %
\section{statistical errors}\label{sect_statistics}
In this section we show how we construct covariance matrices for the polynomial-weighted spectra, compute the Fisher-matrix and derive statistical errors on cosmological parameters, and investigate the signal strength of the weak shear signal, all as a function of the number of polynomials used.

%---  ---%
\subsection{variances of weighted ellipticities}
For deriving the expressions for the signal covariance and the noise covariance needed in forecasting statistical and systematical errors, we derive expressions for the mean and the variance for a weighted set of ellipticities. This derivation is done for a discrete set of weighting coefficients $w_m$ and then generalised to the continuous case, where the weighting is done with a polynomial $p_i(\chi)$. The distribution $p(\epsilon)\dd\epsilon$ of ellipticities $\epsilon$ is assumed to be Gaussian, with a zero mean, variance $\sigma_\epsilon$ and the ellipticities are taken to be intrinsically uncorrelated \citep[i.e. intrinsic alignment effects are discarded, see][for a review]{2009IJMPD..18..173S}.

From a measurement of a set of ellipticities $\epsilon_m$ drawn from the parent distribution $p(\epsilon)\dd\epsilon$ on can estimate the shear by computing the weighted mean $\bar\epsilon$, and the expectation value of the weighted mean $\bra\bar{\epsilon}\ket$ if the drawing of ellipticities is repeated,
\begin{equation}
\bar{\epsilon} = \frac{\sum_m w_m\epsilon_m}{\sum_m w_m}
\rightarrow
\bra\bar{\epsilon}\ket = \frac{\sum_m w_m\bra\epsilon_m\ket}{\sum_m w_m} = 0,
\end{equation}
which vanishes if the mean ellipticity vanishes, $\bra\epsilon_m\ket=0$. For the variance $\bra \bar{\epsilon}^2\ket$ one obtains under the assumption of intrinsicallly uncorrelated ellipticities ,$\bra\epsilon_m\epsilon_n\ket = \sigma_\epsilon^2\delta_{mn}$,
\begin{equation}
\bra\bar{\epsilon}^2\ket = 
\frac{1}{\sum_m w_m\sum_n w_n}\sum_{mn} w_m w_n \bra\epsilon_m\epsilon_n\ket = 
\frac{\sigma_\epsilon^2}{\left(\sum_m w_m\right)^2}\sum_m w_m^2,
\end{equation}
which reduces to the classic Poissonian result if $w_m$ is either 0 or 1,
\begin{equation}
\bra\bar{\epsilon}^2_{ww}\ket = \frac{\sigma_\epsilon^2}{N}
\quad\mathrm{with}\quad
N = \sum_m w_m,
\end{equation}
because $w_m^2=w_m$, and $N$ is defined as the effective number of ellipticities in the sample. The cross-variance for two different sets of weights $w_m$ and $v_n$ is given by:
\begin{equation}
\bra\bar{\epsilon}^2\ket = \frac{\sigma_\epsilon^2}{\sum_m w_m\sum_n v_n}\sum_m w_m v_m.
\end{equation}
In the continuum limit we make the transition
\begin{equation}
\sum_m\ldots\rightarrow\bar{n}\int\dd\chi q(\chi)\ldots
\end{equation}
with the unit normalised galaxy distance distribution $q(\chi)\dd\chi$. The discrete weights $w_m$ and $v_m$ will be replaced by the set of polynomials $p_i(\chi)$ and $p_j(\chi)$. For conserving the normalisation of the weighted galaxy distance distribution $q(\chi)\dd\chi$, we normalise the polynomials
\begin{equation}
p_i(\chi)\leftarrow \frac{p_i(\chi)}{\int\dd\chi\:q(\chi)p_i(\chi)},
\end{equation}
such that the weak shear spectrum $S_{ij}(\ell)$ becomes
\begin{equation}
S_{ij} = 
\bra p_i, p_j\ket = 
\int_0^{\chi_H}\frac{\dd\chi}{\chi^2}\:W_i(\chi) W_j(\chi)\: P(k = \ell/\chi),
\end{equation}
which is diagonal by construction, and the noise covariance,
\begin{equation}
N_{ij} = \frac{\sigma_\epsilon^2}{\bar{n}} \int\dd\chi\:q(\chi)p_i(\chi)p_j(\chi)
\end{equation}
with the shape noise $\sigma_\epsilon$ and the mean density of galaxies $\bar{n}$ per steradian, for which we substitute the numbers $\sigma_\epsilon=0.3$ and $\bar{n} = 40/\mathrm{arcmin}^2$ projected for EUCLID. It can already be seen from the expression for $N_{ij}$ that the weighting of data with high-order polynomials $p_i(\chi)$ will be noisy: $q(\chi)$ is a slowly varying function and the integrals $\int\dd\chi\:q(\chi)\p_i(\chi)$ in the denominator will assume small values if the polynomial is rapidly oscillating. This will ultimately limit the order of the usable polynomials. Again, for $i=j=0$, the standard Poissonian expression $N_{00}=\sigma_\epsilon^2/\bar{n}$ is recovered due to the normalisation of $q(z)$, as well as the weak shear spectrum $S_{00}(\ell) = C_\kappa(\ell)$.

% ---  --- %
\subsection{Fisher-analysis}
The likelihood function for observing Gaussian-distributed modes $\kappa_i(\bmath{\ell})$ of the $p_i(\chi)$-weighted weak lensing convergence for a given parameter set $x_\mu$ is defined as \citep[see][]{1997ApJ...480...22T, 2011arXiv1107.0726C}:
\begin{equation}
\mathcal{L}(\kappa_i(\bmath{\ell})|x_\mu) = 
\frac{1}{\sqrt{(2\pi)^N\mathrm{det}(C)}}
\exp\left(
-\frac{1}{2}\kappa_i(\bmath{\ell})
C^{-1}_{ij}(\bmath{\ell},\bmath{\ell}^\prime)
\kappa_j^*(\bmath{\ell}^\prime)\right)
\end{equation}
with the covariance $C_{ij}(\bmath{\ell},\bmath{\ell}^\prime)\equiv\bra\kappa_i(\bmath{\ell})\kappa_j^*(\bmath{\ell}^\prime)\ket$ which is diagonal in $\ell$ for homogeneous random fields. The $\chi^2$-functional, $\mathcal{L}\propto\exp(-\chi^2/2)$, can be obtained from the logarithmic likelihood and reads:
\begin{equation}
\chi^2 = \sum_\ell\trace\left[\ln C+C^{-1}D\right],
\end{equation}
with the definition of the data matrix $D_{ij} = \kappa_i(\bmath{\ell})\kappa_j(\bmath{\ell})$, using the relation $\ln\mathrm{det}(C) = \trace\ln(C)$ and discarding irrelevant multiplicative prefactors. The second derivatives of the $\chi^2$-functional with respect to cosmological parameters $x_\mu$ evaluated at the point of maximum likelihood yields the Fisher matrix
\begin{equation}
F_{\mu\nu} 
= -\left\bra \frac{\partial^2}{\partial x_\mu\partial x_\nu}\frac{\chi^2}{2}\right\ket
= \sum_\ell \frac{2\ell+1}{2}\trace\left(\frac{\partial}{\partial x_\mu}\ln C\frac{\partial}{\partial x_\nu}\ln C\right)
\end{equation}
with multiplicity $2\ell+1$ because on each angular scale $\ell$ there are $2\ell+1$ statistically independent $m$-modes. The covariance $C_{ij} = S_{ij}+N_{ij}$ can be split up into the signal covariance $S_{ij}$, 
\begin{equation}
S_{ij}(\ell) = 
\left(
\begin{array}{ccc}
S_{00}(\ell) = C_\kappa(\ell) &  & 0 \\
  & \ddots &  \\
0 &  & S_{qq}(\ell)
\end{array}
\right)
\end{equation}
and the noise covariance $N_{ij}$,
\begin{equation}
N_{ij}(\ell) = 
\left(
\begin{array}{ccc}
N_{00}(\ell) = \sigma_\epsilon^2/\bar{n} & \cdots & N_{q0}(\ell) \\
\vdots & \ddots & \vdots \\
N_{0q}(\ell) & \cdots & N_{qq}(\ell)
\end{array}
\right)
\end{equation}
We will work in the limit $\partial S_{ij}/\partial x_\mu\gg\partial N_{ij}/\partial x_\mu$ which is well justified in our case.

\begin{figure}
\resizebox{\hsize}{!}{\includegraphics{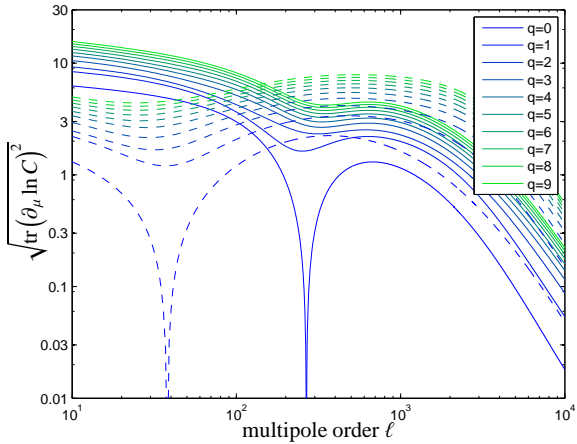}}
\caption{Derivatives $\sqrt{\mathrm{tr}\left(\partial_{\Omega_m}\ln C\right)^2}$ (solid lines) and $\sqrt{\mathrm{tr}\left(\partial_w\ln C\right)^2}$ (dashed lines) as a function of multipole order $\ell$ and cumulative polynomial order $q$. The derivatives have been weighted with the EUCLID covariance.}
\label{fig_derivative}
\end{figure}

Fig.~\ref{fig_derivative} gives an example of how the combination of multiple line of sight-weighted measurements helps to avoid weaknesses in the sensitivity towards cosmological parameters. We plot the sensitivity
\begin{equation}
\sqrt{\trace\left(\frac{\partial\ln C}{\partial x_\mu}\right)^2} = 
\sqrt{\frac{2}{2\ell+1}}\frac{\dd F_{\mu\mu}}{\dd\ell}
\end{equation}
i.e. the ratio between the derivative of the spectrum with respect to a cosmological parameter and the covariance of the measurement, which corresponds to the contributions to the diagonal elements of the Fisher-matrix per $\ell$-mode. The sensitivites of the measurement with respect to the cosmological parameters $\Omega_m$ and $w$ are shown, which exhibit singularities at certain multipoles for a classical non-tomographic measurement. This happens when a parameter affects a certain $\ell$-range with a different sign than others and the spectrum pivots around this value when varying a parameter. Clearly, angular scales in the vicinity of that pivot scale do not add much sensitivity to the Fisher-matrix. The inclusion of a second polynomial, however, avoids this: For $q\geq1$ on, singularities are lifted and the derivatives assume larger values with increasing numbers of polynomials, although this effect saturates for very large numbers of polynomials. At large multipoles, the influence of the shape noise can be observed, which causes the derivatives to drop rapidly, and do not add significant sensitivity to the Fisher matrix on scales larger than $\ell\simeq 3000$.

% ---  --- %
\subsection{statistical errors}
From the Fisher-matrix $F_{\mu\nu}$ one can obtain the Cram{\'e}r-Rao errors,
\begin{equation}
\sigma_\mu^2 = (F^{-1})_{\mu\mu},
\end{equation}
and the 2-dimensional marginalised logarithmic likelihood $\chi^2_\mathrm{m}$ around the fiducial model $x_\mu^*$,
\begin{equation}
\chi^2_\mathrm{m} = 
\left(
\begin{array}{c}
x_\mu-x_\mu^* \\ x_\nu - x_\nu^*
\end{array}
\right)^t
\left(
\begin{array}{cc}
(F^{-1})_{\mu\mu} & (F^{-1})_{\mu\nu} \\
(F^{-1})_{\nu\mu} & (F^{-1})_{\nu\nu}
\end{array}
\right)^{-1}
\left(
\begin{array}{c}
x_\mu - x_\mu^* \\ x_\nu - x_\nu^*
\end{array}
\right).
\end{equation}

\begin{figure}
\resizebox{\hsize}{!}{\includegraphics{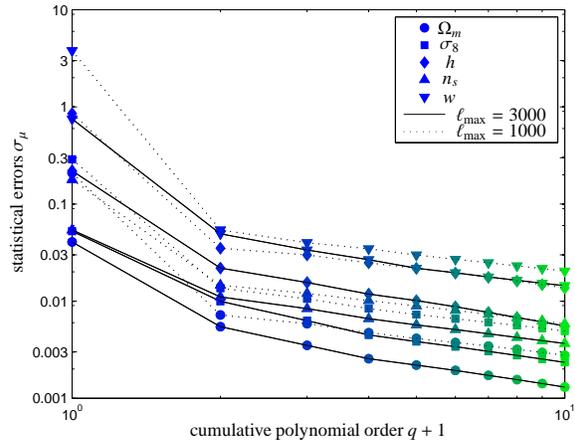}}
\caption{Statistical errors on the estimates of the cosmological parameters $\Omega_m$ (circles), $\sigma_8$ (squares), $h$ (lozenges), $n_s$ (triangles, pointing up) and $w$ (triangles, pointing down) resulting as Cram{\'e}r-Rao errors from computing the Fisher-matrix for the weak lensing spectrum $S_{ii}(\ell)$, as a function of cumulative polynomial order $q$, for EUCLID survey characteristics. The maximum multipole consideres is $\ell_\mathrm{max}=1000$ (dotted lines) and $\ell_\mathrm{max}=3000$ (solid lines).}
\label{fig_errors}
\end{figure}

Fig.~\ref{fig_errors} summarises statistical errors on cosmological parameters resulting from the Fisher-matrix analysis, cumulative in $q$ and for $\ell=1000$ as well as for $\ell=3000$. As expected, statistical errors drop with larger number of polynomials and are smaller if more multipoles are considered. A very fascinating feature of the plot is the approximate scaling of the error with the inverse root of the number of polynomials, $\sigma\propto 1/\sqrt{q}$, which one expects from Poissonian arguments because each spectrum $S_{ii}(\ell)$ adds statistically independent information.

\begin{figure*}
\vspace{0.5cm}
\resizebox{0.85\hsize}{!}{\includegraphics{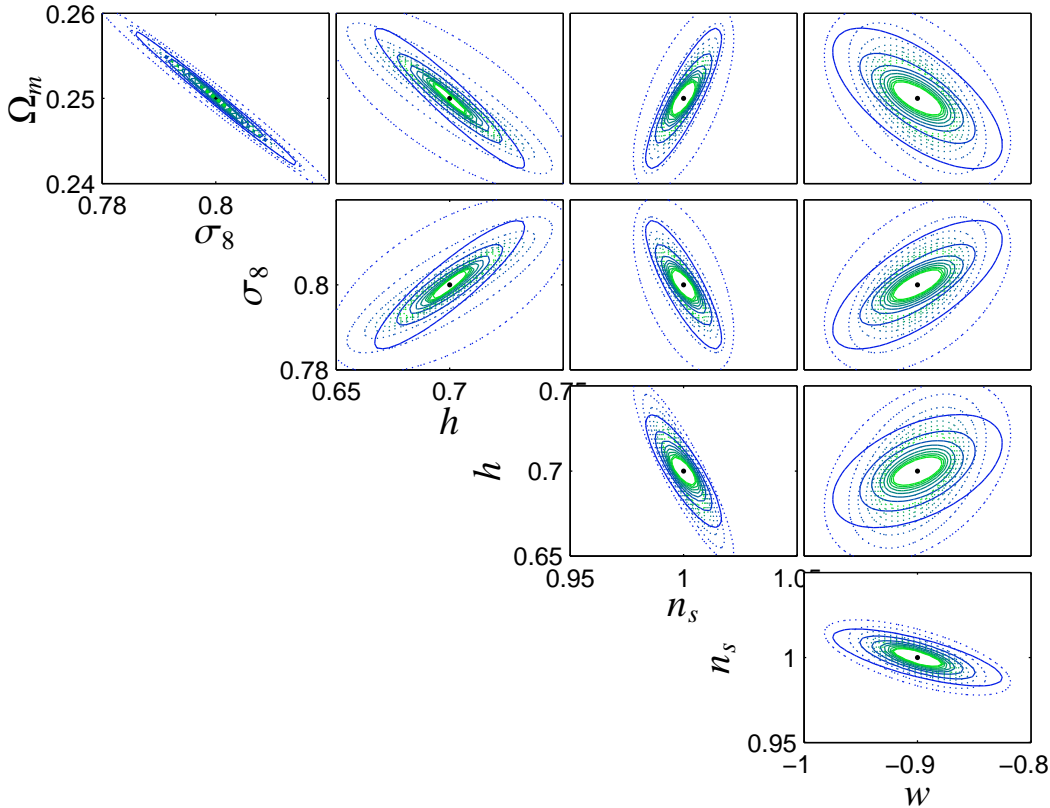}}
\vspace{0.5cm}
\caption{Constraints on the cosmological parameters $\Omega_m$, $\sigma_8$, $h$, $n_s$ and $w$ from EUCLID using tomography with orthogonal polynomials. The ellipses mark $1\sigma$ confidence regions and decrease in size with increasing cumulative polynomial order ($q=0$ in blue to $q=9$ in green), for $\ell_\mathrm{max}=1000$ (dotted lines) and $\ell_\mathrm{max}=3000$ (solid lines).}
\label{fig_fisher}
\end{figure*}

\begin{figure}
\resizebox{\hsize}{!}{\includegraphics{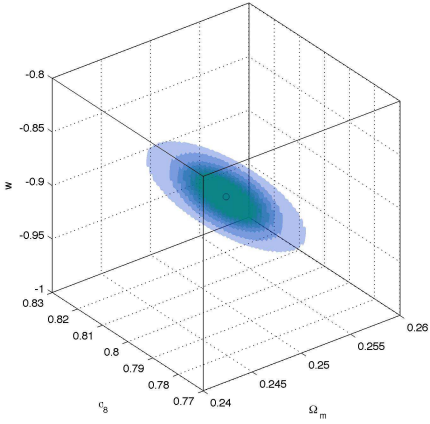}}
\caption{$1\sigma$-ellipsoids in the space spanned by $(\Omega_m,\sigma_8,w)$, cumulative in polynomial order $q$, from $q=2$ (largest ellipsoid) up to $q=9$ (smallest ellipsoid) with the maximum multipole order set to $\ell=3000$. Observational characteristics correspond to those of the EUCLID mission.}
\label{fig_ellipsoid}
\end{figure}

2-dimensional marginalised likelihoods for all parameter pairs are shown in Fig.~\ref{fig_fisher}, for $\ell=1000$ and $\ell=3000$ as the maximum multipole considered. In the confidence contours, up to 10 polynomials were combined. Clearly, the uncertainty in cosmological parameters decreases with larger numbers of polynomials used, as well as for an increased multipole range. Additionally, the parameter degeneracies change a little when more polynomials are used, which is caused by differing sensitivities of the lensing signal with distance. A 3-dimensional view of the marginalised likelihood of the parameters $\Omega_m$, $\sigma_8$ and $w$ is given in Fig.~\ref{fig_ellipsoid}. It illustrates nicely the nested $1\sigma$-ellipsoids which become smaller for larger number of polynomials combined in the measurement. Comparing the performance of the TaRDiS-polynomials to other tomographic methods shows that they operate on very similar levels of performance, perhaps with a small advantage for the cosmological parameters $\Omega_m$ and $w$. This is due to the mechanism that the values in the Fisher-matrix are maximised if the covariance becomes diagonal due to a match between the true cosmology and the assumed cosmology used for constructing the polynomials. This should be taken with a grain of salt, however, because part of the statistical error would be transported to the systematical error budget. The impact of starting off with a wrong cosmological model for construction of the TaRDiS-polynomials on the estimation of parameters will be the topic of the next chapter.

% ---  --- %
\subsection{signal to noise-ratio} 
\begin{figure}
\resizebox{\hsize}{!}{\includegraphics{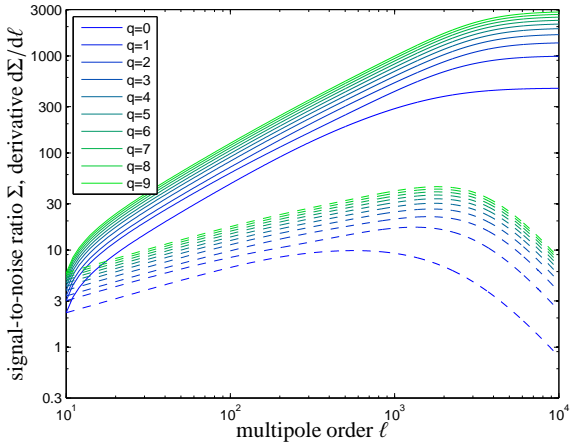}}
\caption{Signal to noise-ratio $\Sigma(\ell)$ (solid lines) and the contribution $\dd\Sigma/\dd\ell$ by each multipole (dashed lines) as a function of inverse angular scale $\ell$ and cumulative polynomial order $q$. The shape noise corresponds to that of the projected EUCLID performance.}
\label{fig_s2n}
\end{figure}

Analogous to the definition of the Fisher-matrix we construct the cumulative signal to noise-ratio $\Sigma$,
\begin{equation}
\Sigma^2 = 
\sum_\ell\frac{2\ell+1}{2}\trace\left(C^{-1}SC^{-1}S\right) = 
\sum_\ell\frac{2\ell+1}{2}\trace\left(C^{-1}S\right)^2,
\label{eqn_s2n}
\end{equation}
from the covariance matrices with the multiplicity $2\ell+1$. The cumulative signal to noise-ratio $\Sigma$ and the differential contribution $\dd\Sigma/\dd\ell$ of each multipole is summarised in Fig.~\ref{fig_s2n}. Clearly, increasing $q$ or $\ell$ increases the signal up to multipole orders of a few thousand. The growth of $\dd\Sigma(\ell)/\dd\ell\propto\sqrt{2\ell+1}$ is driven by the reduced cosmic variance until the shape noise limits the measurability of the spectra. Consequently, the integrated signal to noise ratio settles off at a few hundred, and adding statistically independent information by using new polynomials increases the signal strength by almost an order of magnitude and reaches values comparable to the primary CMB temperature anisotropy spectrum \citep{2002PhRvD..65b3003H}. These numbers correspond well to those derived by \citet{2009MNRAS.395.2065T} if one works in the approximation of a Gaussian covariance - non-Gaussian contributions can significantly lower the signal strength. This can be expected from eqn.~(\ref{eqn_s2n}), as the signal to noise ratio is invariant under orthogonal transformations diagonalising either $S_{ij}$ or $N_{ij}$. 

% --- section:  --- %
\section{systematical errors}\label{sect_systematics}
Naturally, the set of polynomials used for analysing the data needs to be constructed for specific cosmology, so the question arises if an imprecise prior knowledge of the cosmological model has an impact on the parameter estimates from the weighted weak lensing spectra. Any incompleteness or imperfection in the model used for interpreting the data is going to shift the estimated parameter values away from their true values and introduces parameter estimation biases. Specifically, we consider three cases: 

Firstly, a time-varying equation of state ($w=-0.8+0.2(1-a)$) when in reality the equation of state of dark energy is constant ($w=-0.9$), i.e. a systematic which is not a degree of freedom of the model, and secondly wrongly assumed $\Omega_m$- and $\sigma_8$-values (0.3 instead of 0.25 and 0.85 instead of 0.8, respectively) as a strong systematic. In addition, we compute biases in the estimation of cosmological parameters if the redshift distribution of galaxies substituted in the polynomial construction is not the true one.

% ---  --- %
\subsection{parameter estimation bias}
For the Gaussian likelihood function $\mathcal{L}\propto\exp(-\chi^2/2)$ for a parabolic $\chi^2$-functional one can identify $\chi^2\equiv\sum_\ell\trace\left[\ln C + C^{-1}D\right]$ with the covariance $C_{ij}$ and the data matrix $D_{ij}$. Now, the fit of the true model $C_t$ to the data would give rise to the correct $\chi^2_t$-functional, 
\begin{equation}
\chi^2_t=\sum_\ell\trace\left[\ln C_t + C^{-1}_tD\right], 
\end{equation}
whereas the assumption of a wrong model yields 
\begin{equation}
\chi^2_f=\sum_\ell\trace\left[\ln C_f + C^{-1}_fD\right], 
\end{equation}
which in general will assume its global minimum at parameter values different than $\chi^2_t$. The distance $\bmath{\delta}$ between the best fit values $\bmath{x}_t$ of the true model and $\bmath{x}_f$ of the false model will then be the parameter estimation bias. This parameter estimation bias can be quantified with a second-order Taylor expansion of the $\chi^2_f$-functional for the wrong model around the best-fit point $\bmath{x}_t$ of the true model \citep[see][]{2007MNRAS.381.1347C},
\begin{equation}
\chi_f^2(\bmath{x}_f) = 
\chi_f^2(\bmath{x}_t) + 
\sum_\mu\frac{\partial}{\partial x_\mu}\chi_f^2(\bmath{x}_t)\: \delta_\mu + 
\frac{1}{2}\sum_{\mu,\nu}\frac{\partial^2}{\partial x_\mu\partial x_\nu}\chi_f^2(\bmath{x}_t)\: \delta_\mu\delta_\nu,
\end{equation}
where the parameter estimation bias vector $\bdelta \equiv \bmath{x}_f-\bmath{x}_t$. The best-fit position $\bmath{x}_f$ of $\chi_f^2$ can be recovered by extremisation of the ensemble-averaged $\bra\chi_f^2\ket$, yielding
\begin{equation}
\underbrace{\left\bra\frac{\partial}{\partial x_\mu}\chi_f^2\right\ket_{\bmath{x}_t}}_{\equiv a_\mu} = 
\sum_\nu\underbrace{-\left\bra\frac{\partial^2}{\partial x_\mu\partial x_\nu}\chi_f^2\right\ket_{\bmath{x}_t}}_{\equiv G_{\mu\nu}}\delta_\nu,
\end{equation}
which is a linear system of equations of the form 
\begin{equation}
\sum_\nu G_{\mu\nu}\delta_\nu = a_\mu\rightarrow \delta_\mu = \sum_\nu (G^{-1})_{\mu\nu}a_\nu,
\end{equation}
where the two quantities $G_{\mu\nu}$ and $a_\mu$ follow from the derivatives of the averaged $\chi_f^2$-functional, evaluated at $\bmath{x}_t$, with $\bra D\ket = C_t$ and the multiplicity $2\ell+1$ added for each $\ell$-mode:
\begin{equation}
a_\mu = \sum_\ell\frac{2\ell+1}{2}
\trace\left[\frac{\partial}{\partial x_\mu}\ln C_f\left(\id-C_f^{-1}C_t\right)\right],
\end{equation}
This vector reduces to $a_\mu=0$ if $C_t=C_f$ ($\mathrm{id}$ being the identity matrix). Furthermore, 
\begin{eqnarray}
G_{\mu\nu} 
& = & \sum_\ell\frac{2\ell+1}{2}
\trace
\left[C_f^{-1}\frac{\partial^2}{\partial x_\mu\partial x_\nu}C_f\left(C_f^{-1}C_t-\id\right)\right]
\nonumber\\ 
& - & \sum_\ell\frac{2\ell+1}{2}
\trace
\left[\frac{\partial}{\partial x_\mu}\ln C_f\frac{\partial}{\partial x_\nu}\ln C_f\left(2C_f^{-1}C_t-\id\right)\right]
\end{eqnarray}
which simplifies to $G_{\mu\nu}=F_{\mu\nu}$ in the case of choosing the correct model, such that the parameter estimation bias vanishes. The same happens for $q=0$, i.e. if only the polynomial $p_0(\chi)\equiv 1$ is used. In the derivation outlined above, the identities
\begin{equation}
\frac{\partial}{\partial x_\mu}\ln C = C^{-1}\frac{\partial}{\partial x_\mu}C
\quad\mathrm{and}\quad
\frac{\partial}{\partial x_\mu}C^{-1} = - C^{-1}\left(\frac{\partial}{\partial x_\mu}C\right)C^{-1}
\end{equation}
were used. This formalism is a generalisation of the case of diagonal covariance matrices \citep{2007MNRAS.381.1347C, 2008MNRAS.391..228A, 2009MNRAS.392.1153T, 2011MNRAS.tmp..612M}, for which it has been shown to work well by comparison with results from Monte-Carlo Markov chains \citep{2010MNRAS.404.1197T}. We need to employ this more general formalism because the signal covariance $S_{ij}(\ell)$ ceases to be diagonal if the cosmology used for constructing the polynomials differs from the true cosmology. Other examples of non-diagonal covariances include 3-dimensional cosmic shear \citep{2003MNRAS.343.1327H, 2008MNRAS.389..173K}. It should be emphasised that we are only investigating the impact of systematics or a wrongly chosen initial cosmology on the construction of polynomials and that systematics control plays a very important role in parameter estimation from weak lensing \citep[among others, see][]{2002A&A...396..411K, 2006MNRAS.366..101H, 2007NJPh....9..444B, 2011arXiv1105.1075S}, which we are not touching here.

% ---  --- %
\subsection{systematical errors}
\begin{figure*}
\vspace{0.5cm}
\resizebox{0.85\hsize}{!}{\includegraphics{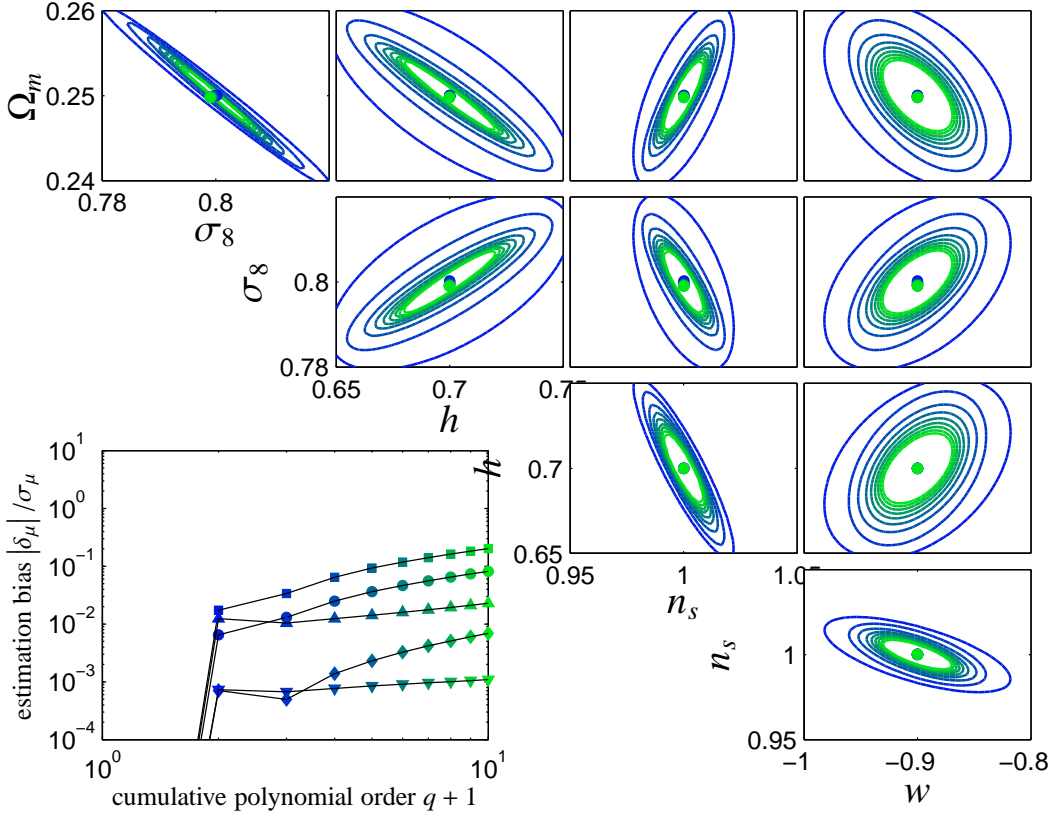}}
\vspace{0.5cm}
\caption{Parameter estimation biases $(\delta_\mu,\delta_\nu)$ in the parameters $\Omega_m$, $\sigma_8$, $h$, $n_s$ and $w$, superimposed on the $1\sigma$-confidence regions, if the polynomials $p_i(\chi)$ have been constructed for $w$CDM with an evolving equation of state parameterised by $w_0=-0.8$ and $w_a=-0.2$ instead of $w$CDM with a constant equation of state with $w=-0.9$. In the estimation biases, the dot colour is proportional to the cumulative order $q$ of the polynomials, and is plotted again as a function of $q$, for $\Omega_m$ (dots), $\sigma_8$ (squares), $h$ (lozenges), $n_s$ (triangles, pointing up), $w$ (triangles, pointing down). For all computations, the EUCLID survey characteristics were used, and the data was accumulated up to $\ell_\mathrm{max}=1000$.}
\label{fig_bias_1}
\end{figure*}

We constructed TaRDiS-polynomials for three wrong assumptions: As the first example, we consider a rather small change in the dark energy model, namely a time varying gequation of state instead of a constant one. These two cosmologies have the same average dark energy equation of state, but the degree of freedom of a time varying $w$ is contained in the true cosmology. For this case, Fig.~\ref{fig_bias_1} shows the estimation bias in the true dark energy model, when the model used for constructing the polynomials had been a time varying equation of state with the same average eos-parameter. The figure illustrates that such a mistake has a minor impact on the estimation of parameters, as biases are smaller compared to the statistical precision by at least an order of magnitude. Most estimation bias can be found in the normalisation $\sigma_8$.

\begin{figure*}
\vspace{0.5cm}
\resizebox{0.85\hsize}{!}{\includegraphics{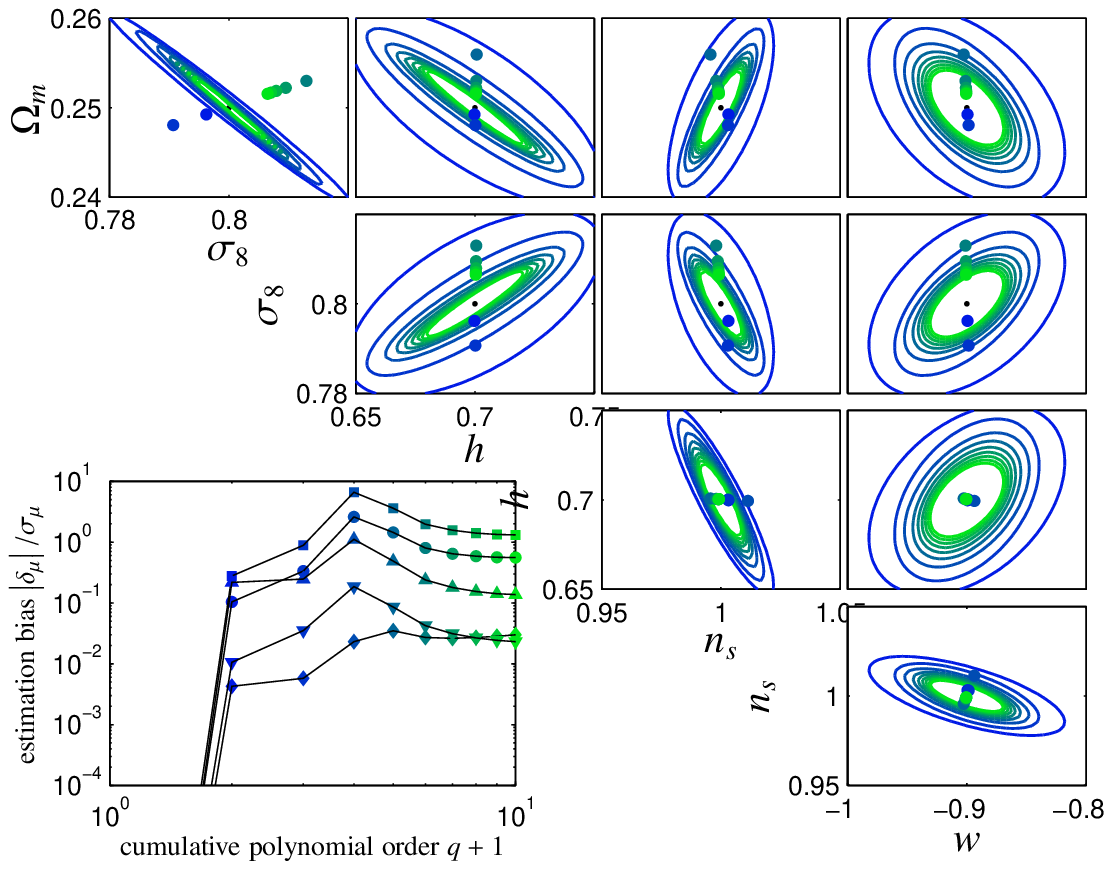}}
\vspace{0.5cm}
\caption{Parameter estimation biases $(\delta_\mu,\delta_\nu)$ in the cosmological parameters $\Omega_m$, $\sigma_8$, $h$, $n_s$ and $w$, superimposed on the $1\sigma$-confidence regions, if in the construction of the polynomials $\Omega_m$ and $\sigma_8$ were set too high (0.275 and 0.85 instead of 0.25 and 0.8, respectively) relative to the fiducial cosmology. The colour of the dots indicates the cumulative polynomial order $q$, for $\Omega_m$ (dots), $\sigma_8$ (squares), $h$ (lozenges), $n_s$ (triangles, pointing up) and  $w$ (triangles, pointing down). For the plot, EUCLID survey characteristics were assumed, and the signal integrated up to $\ell_\mathrm{max}=1000$.}
\label{fig_bias_2}
\end{figure*}

Biases in the estimation of cosmological parameters if for the construction of the polynomials the wrong values for $\Omega_m$ and $\sigma_8$ have been used, are summarised in Fig.~\ref{fig_bias_2}. This case is a much stronger systematic compared to the previous case. There are strong biases in particular in $\Omega_m$ and $\sigma_8$, and to a lesser extend in $n_s$, whereas $h$ and $w$ are not strongly affected. Furthermore, the biases in $\Omega_m$ and $\sigma_8$ are in a direction almost orthogonal to the orientation of the degeneracy, indicating a patricularly strong impact.These estimation biases, however, can be reduced by including a larger number of polynomials. In that way, a reduction of the bias to values similar to the statistical error is possible.

\begin{figure*}
\vspace{0.5cm}
\resizebox{0.85\hsize}{!}{\includegraphics{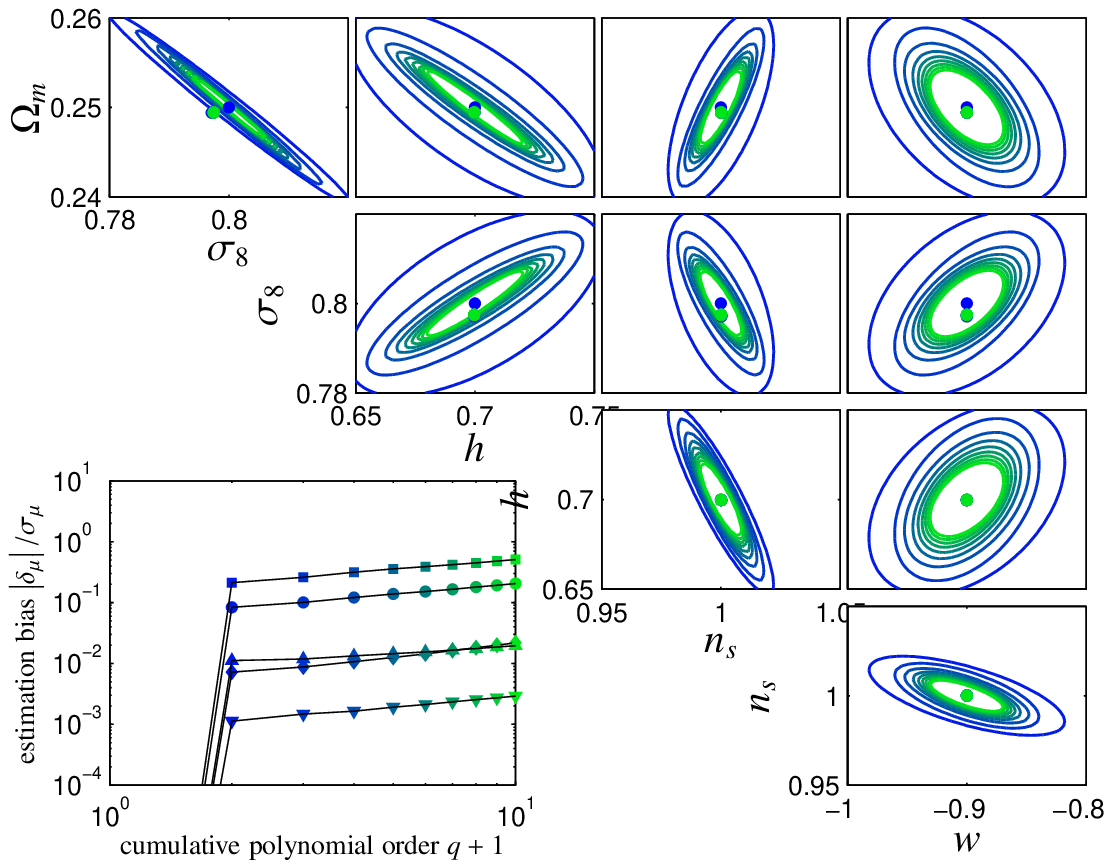}}
\vspace{0.5cm}
\caption{Biases $(\delta_\mu,\delta_\nu)$ in the cosmological parameters $\Omega_m$, $\sigma_8$, $h$, $n_s$ and $w$, superimposed on the $1\sigma$-confidence regions, if the true redshift distribution of galaxies differs from the observed one by a convolution with a Gaussian with $\sigma_z=0.05$. The dot colour changes with cumulative polynomial order $q$, and is replotted as a function of $q$ in the inset for $\Omega_m$ (dots), $\sigma_8$ (squares), $h$ (lozenges), $n_s$ (triangles, pointing up) and $w$ (triangles, poiting down). The errors and biases correspond to the EUCLID survey up to the multipole order $\ell_\mathrm{max}=1000$.}
\label{fig_bias_3}
\end{figure*}

Finally, biases due to using a convolved redshift distribution in the polynomial construction whereas the signal is in reality generated by the unconvolved redshift distribution have been computed in Fig.~\ref{fig_bias_3}. As a very simple model for the error in the measurement of the galaxy redshift distribution $q(z)\dd z$ we used the convolution with a Gaussian \citep{2006ApJ...636...21M},
\begin{equation}
r(z) = \frac{1}{\sqrt{2\pi\sigma_z^2}}\exp\left(-\frac{z^2}{2\sigma_z^2}\right)
\quad\mathrm{with}\quad
\sigma_z = 0.05,
\end{equation}
which is of course too coarse to model errors in a proper weak lensing measurement, but will serve as an example. In contrast to the previous example, systematical errors in the parameters increase with larger numbers of polynomials used, and reach values of the order of the statistical error, in particular for $\Omega_m$ and $\sigma_8$, and to a lesser extend $h$. Quite similarly, the direction of the bias move the best fit point quickly away from the true cosmology because it is at roughly right angles relative to the degeneracy direction.

Given the magnitude of systematical errors in comparison to their statistical accuracies, parameter estimation biases due to a wrongly chosen cosmology appear unlikely to affect measurements. In particular, if values estimated from unweighted tomography are used, the cosmological parameters are known well enough that TaRDiS-polynomials can be constructed in a reliable way and consequent parameter estimation biases would always be subdominant in relation to the statistical errors. In addition, it is always possible to iterate between parameter estimation and polynomial construction for narrowing down estimation biases.

% ---  --- %
\subsection{cross validation}
\begin{figure}
\resizebox{\hsize}{!}{\includegraphics{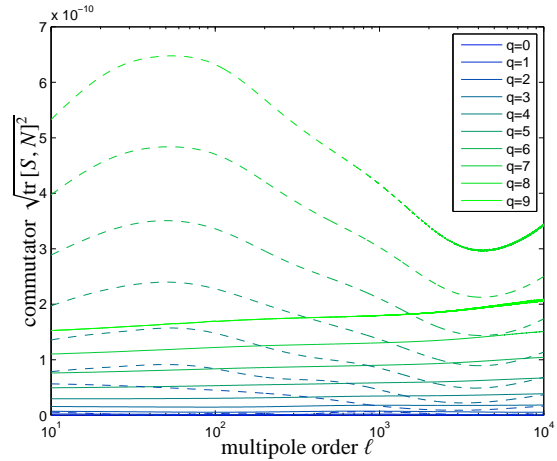}}
\caption{Commutator $\sqrt{\trace\left[S,N\right]^2}$ between the signal covariance $S_{ij}$ and the noise covariance $N_{ij}$, cumulative in polynomial order $q$ (indicated by colour), for the two systematics considered here: assuming a time-varying instead of a constant $w$CDM model ($w_0=-0.8$ and $w_a=-0.2$, solid lines) and choosing a too high matter density and a too high normalisation of $P(k)$ ($\Omega_m=0.275$ and $\sigma_8=0.85$, dashed lines), both for EUCLID.}
\label{fig_commutators}
\end{figure}

A possible way of validating the correctness of the assumed cosmology for the construction of the orthogonal set of polynomials would be to estimate the cross spectra $C_{ij}(\ell)$ which should be compatible with zero for the correct choice \citep[similarly to][]{2005PhRvD..72d3002H}, or alternatively, one can take advantage of the non-commutativity $\left[S,N\right]=SN-NS\neq 0$ of the signal covariance $S$ and the noise covariance $N$. If the polynomials have been constructed for the true cosmology with the first normalisation variant eqn.~(\ref{eqn_norm_scalar}), $S$ for unit-normalised polynomials is equal to the unit matrix and would always commute with $N$. In the case of the wrong cosmology, $S$ and $N$ are symmetric matrices with different eigensystems, and $\left[S,N\right]$ does not vanish. In Fig.~\ref{fig_commutators} this commutator is given as a function of multipole order $\ell$ and depending on the number of polynomials used. For giving a single number quantifying the non-commutativity we chose to compute the trace $\trace\left[S,N\right]^2$, because $\trace\left[S,N\right]$ without squaring always vanishes. The stronger systematic (detuning $\Omega_m$ and $\sigma_8$ from their fiducial values) generates larger values compared to the wrong assumption of a time-varying equation of state of the dark energy component. Naturally, when using a single polynomial $q=0$ the commutator is always zero.

Given the minor impact on choosing a wrong cosmological impact on parameter estimation from weak lensing spectra weighted with orthogonal polynomials, we point out that it should be possible to employ the polynomials iteratively by alternating between parameter estimation and polynomial construction. At the same time we emphasise the possible usefulness of measuring cross spectra $C_{ij}(\ell)$, $i\neq j$, or the commutator $\trace\left[S,N\right]^2$ for validating the cosmological model used for data analysis.

% --- section: summary --- %
\section{Summary}\label{sect_summary}
The topic of this paper is a novel method of carrying out tomographic weak lensing measurements by line of sight-weighting of the weak lensing convergence with specifically constructed orthogonal polynomials.
\begin{enumerate}
\item{The TaRDiS-polynomials have been constructed with a Gram-Schmidt orthogonalisation procedure in order to diagonalise the signal covariance matrix, and to project out statistically independent information of the weak lensing field by performing weak lensing tomography. They differ from traditional tomography and 3-dimensional weak lensing methods in the respect that the signal covariance is diagonal instead of the noise covariance.}
\item{The statistical error forcasts for cosmological parameters was investigated with the Fisher-matrix formalism. Because of the polynomial's property of providing statistically independent information of the weak convergence field, one sees a particularly simple square-root scaling of the signal to noise-ratio and the statistical errors with the maximum multipole considered and the number of polynomials used. Clearly, statistical errors can be reduced when combining polynomials, where a small advantage can be observed in comparison to traditional tomography. This is due to the fact that in case of a matching between the assumed and the true cosmology, the Fisher-matrix assumes the largest possible values because of teh diagonalised covariance and provides consequently the smallest statistical errors.}
\item{We extended the Fisher-matrix formalism for investigating how the assumption of a wrong cosmology in the construction of the polynomials impacts on the estimation of cosmological parameters and to what extend biases are introduced. We assumed three systematically wrong priors: assumtion of a time varying instead of a constant dark energy equation of state parameter, as an example of a degree of freedom not contained in the model, a wrong $\Omega_m$- and $\sigma_8$-pair, as being the two most prominent parameters determining the strength of the weak lensing signal, and finally a convolved galaxy redshift distribution. All three systematics had a minor impact on the parameter estimation, with the biases decreasing if a larger number of polynomials was used. With 10 polynomials biases were of the order of the statistical error even for strong mismatches in the choice of the initial cosmology. Only the assumption of a wrong redshift distribution generates roughly constant biases, which, when normalised to the statistical errors, increase with the number of polynomials used. Our extension of the Fisher-matrix formalism treats the parameter estimation biases in full generality and does not assume a diagonal shape of neither the signal nor the noise covariance.}
\item{The accuracy needed for constructing viable polynomials corresponds to the statistical error reachable with a simple unweighted convergence spectrum, which is readily available, and can be improved by adding additional priors in the form of the CMB-likelihood, parameter constraints form baryon acoustic oscillations or from supernovae.}
\item{Additionally, it is possible to compute diagnostics for the choice of the prior cosmological model used for constructing the cosmology: Examples include the cross-spectra $S_{ij}(\ell)$ for two different polynomials $i\neq j$, which should vanish for correctly-constructed polynomials, and the commutator $\left[S,N\right] = SN-NS$ between the signal and the noise covariance, which should vanish if the signal covariance is equal to the unit matrix for a certain choice of normalisation for the polynomials.}
\end{enumerate}

We plan to extend our research on the usage of orthogonal polynomials to the weak lensing bispectrum in a future paper, and to carry out forecasts on dark energy cosmologies from combined constraints with spectrum and bispectrum tomography.

% --- section: acknowledgements --- %
\section*{Acknowledgements}
We would like to thank Matthias Bartelmann, David J. Bacon and Ramesh Narayan for their suggestions and very helpful comments, and Philipp M. Merkel for suggesting the usage of $\trace\left[S,N\right]^2$ in Fig.~\ref{fig_commutators}. We acknowledge the use of the fantastic icosmo-resource \citep{2011A&A...528A..33R} for comparing performances. BMS's work was supported by the German Research Foundation (DFG) within the framework of the excellence initiative through the Heidelberg Graduate School of Fundamental Physics. LH receives funding from the Swiss science foundation.

% --- section: SZ definitions --- %
\bibliography{bibtex/aamnem,bibtex/references}
\bibliographystyle{mn2e}

% --- section appendix --- %
\appendix

\section{weighted lensing spectra}
\begin{figure}
\resizebox{\hsize}{!}{\includegraphics{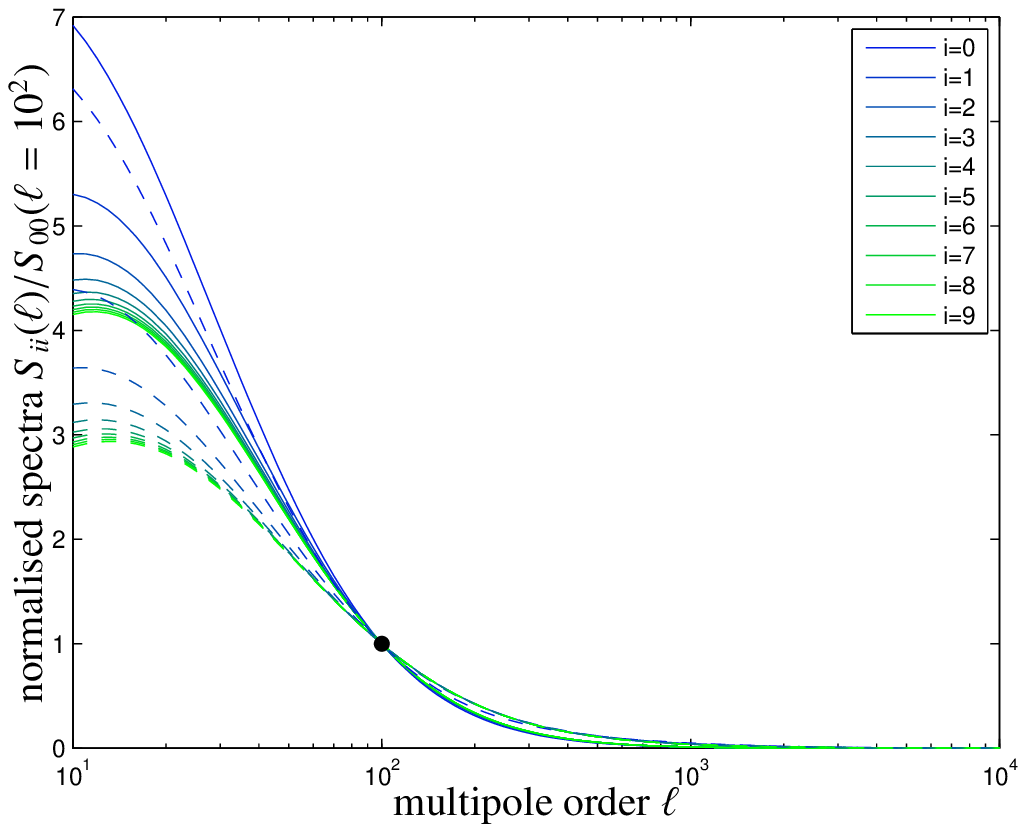}}
\caption{$p_i(\chi)$-weighted weak lensing spectra $S_{ii}(\ell)$, normalised to the value $S_{00}(\ell)\equiv C_\kappa(\ell)$ at $\ell=10^2$ (indicated by the black dot) for illustrating the change in shape introduced by the polynomials. The weighting polynomials have been constructed for a linear (solid line) and a nonlinear (dashed line) CDM spectrum.}
\label{fig_relative_spectra}
\end{figure}

Fig.~\ref{fig_relative_spectra} provides an alternative representation for Fig.~\ref{fig_spectra} of the polynomial-weighted weak lensing spectra $S_{ii}(\ell)$. Here, the spectra are normalised to the value $S_{00}(\ell=10^2)$, such that the difference is shape is easier to see. In particular the low multipoles are suppressed in amplitude by the weighting with $p_i(\chi)$.

\bsp

\label{lastpage}

\end{document}